\documentclass{emulateapj}

\slugcomment{Accepted for publication in \apj}

\shorttitle{XMM-Newton Observations of Galaxy Groups}
\shortauthors{Voevodkin et al.}

\def\head{ \vbox to 0pt{\vss \hbox to 0pt{\hskip 440pt\rm
      LA-UR-07-0764\hss} \vskip 25pt}}

\begin{document}

\head

\title{X-ray Observations of Optically Selected Giant
  Elliptical-Dominated Galaxy Groups}

\author{
Alexey Voevodkin,\altaffilmark{1,5}
Christopher J. Miller,\altaffilmark{2}
Konstantin Borozdin,\altaffilmark{1}
Katrin Heitmann,\altaffilmark{1}
Salman Habib,\altaffilmark{1}
Paul Ricker,\altaffilmark{3}
and Robert C. Nichol\altaffilmark{4}}

\altaffiltext{1}{Los Alamos National Laboratory, Los Alamos, NM, USA}
\altaffiltext{2}{
NOAO/Cerro Tololo Interamerican Observatory
La Serena, Chile}
\altaffiltext{3}{Department of Physics and Astronomy, University of Illinois,
Urbana-Champaign, IL, USA}
\altaffiltext{4}{Institute of Cosmology and Gravitation,
University of Portsmouth,
Portsmouth, PO1 2EG, UK}
\altaffiltext{5}{Space Research Institute, Moscow, Russia}

\begin{abstract}
  We present a combined optical and X-ray analysis of three optically
  selected X-ray bright groups with giant elliptical galaxies in the
  center.  These massive ellipticals were targeted for
  \emph{XMM-Newton} X-ray observations based on their large velocity
  dispersions and their proximity to a nearby \emph{ROSAT} X-ray
  source.  Additionally, these targets are significantly brighter in
  the optical than their nearest neighbors.  We show that one of these
  systems meets the standard criteria for a fossil group. While the
  other two systems have a prominent magnitude gap in the E/S0
  ridgeline, they do not appear to have reached the fossil-like final
  stage of group evolution.
\end{abstract}


\section{Introduction}

X-ray bright giant elliptical galaxies have gained considerable
attention recently as possible end points in the evolution of galaxy
groups.  It is often suggested that these systems form in the process
of merging of smaller galaxies with the associated loss of the
progenitors' disk component and the formation of a common, group-sized
dark matter halo.  Details of this process are still unclear, and the
formation process is sensitive to both the cosmological parameters and
astrophysical processes involved.

The possibility that members of compact groups could merge to form a
large elliptical galaxy on timescales much shorter than the Hubble
time was first predicted from early numerical simulations
~\cite{1989Natur.338..123B}.  In 1993, Ponman \& Bertram suggested
that a large elliptical galaxy which formed through mergers of smaller
galaxies could retain its diffuse halo. Such an object would exhibit a
high X-ray luminosity, while its optical light would be dominated by a
single massive galaxy. In 1994, Ponman et al.  reported the discovery
of just such an object and termed it a ``fossil group'' (FG). Other
researchers have discovered similar objects and have employed terms
such as over-luminous elliptical galaxies (OLEGs,
~\cite{1999ApJ...520L...1V}). The number of these systems reported in
the literature is growing. Similar to the early days of galaxy cluster
research, the precise object definitions used by researchers show
large variations ~\cite{1999ApJ...520L...1V,
  1999ApJ...514..133M,2000ApJS..126..209R, 2003MNRAS.343..627J,
  2004ApJ...612..805S,2004AdSpR..34.2525Y, 2005ApJ...624..124U}.  In
general, the search criteria usually include a cut for both absolute
optical magnitude of the central object, and extended X-ray emission,
as well as the requirement for an optical magnitude gap between the
first and the second brightest member of the group.

While both observations and numerical N-body simulations firmly
establish the existence of fossil-group-like objects, there are
several remaining questions to be addressed.  The theory of
hierarchical structure formation predicts that more massive objects
form by merging of less massive objects, and consequently, that mass
functions for structures of different mass, such as clusters and
groups of galaxies, should be self-similar. While the expected
self-similarity is seen in N-body simulations, it seems to break down
observationally.  The mass function of clusters is in general
agreement with simulations, but even the most massive groups
demonstrate a lack of low-mass satellites and exhibit mass functions
similar to the Local Group ~\cite{2007arXiv0704.2604D}. These
conclusions should be considered preliminary, as they have been so far
based on a small number of studied systems.

Two theories have been proposed as models for the formation of giant
ellipticals: {\it evolutionary formation through merging} and {\it in
  situ through accretion}.  The first scenario suggests that they are
the end-point of group/cluster evolution through mergers and dynamical
friction. The second suggests that they are the massive-end point of
the elliptical galaxy distribution or that they formed initially with
a deficit of small galaxies
~\cite{1999ApJ...514..133M,2003MNRAS.343..627J,2004AdSpR..34.2525Y,2006AJ....131..158M}.
Both of these theories have their own difficulties when trying to
explain the observations.  If these systems have evolved from groups
or clusters one would expect to see cool cores, but observationally
cool cores are not seen at all or are smaller than expected
~\cite{2004MNRAS.349.1240K,2006MNRAS.369.1211K,2004ApJ...612..805S}.
Mass-to-light ratios of fossil groups are unusually high (Khosroshahi
et al.  2007), suggesting that the star-formation history of their
member galaxies is different from other groups and clusters.  They
also appear to have higher than expected concentrations as measured
from their density profiles.

The fraction of fossil-like groups among the general population of
groups may well be a sensitive probe of structure formation. While
both N-body simulations and observations predict that fossil groups
represent a significant fraction of all groups in the mass range
~$10^{13}\,M_\odot$-$10^{14}\,M_\odot$, existing observational
statistics do not provide tight constraints for comparison.
Historically, fossil groups were selected from X-ray surveys. Due to
to the relatively spotty or shallow sky coverage of these X-ray
observations, the number of identified fossil groups is low. Vikhlinin
et al. (1999) estimated that OLEGs comprise ~20\% of all groups and
clusters of comparable luminosity, but this estimate was based on only
4 objects found in a $Rosat$ survey of extended objects. Jones et al.
(2003) found a sample of 6 fossil groups satisfying their criteria and
calculated their fraction to be 8-20\%. Altogether fewer than 20
objects have been studied so far~\cite{2006AJ....131..158M}, and
temperature measurements are available only for a
handful~\cite{2007MNRAS.377..595K}.  In addition, the existing sample
is somewhat heterogeneous as various selection criteria were used in
different studies.  Recently fossil group candidates were identified
in the \emph{Sloan Digital Sky Survey} (SDSS) data
\cite{2007AJ....134.1551S}.  Between 6 and 34 candidates satisfying
their criteria were found, depending on the minimum redshift range and
radius of search for group members.

Recent and planned optical surveys provide an opportunity to expand
the sample of fossil group candidates based on optical selection of
giant ellipticals. We report here the results of our attempt for such
an optical selection (see also Mulchaey \& Zabludoff~1999, Santos et
al.~ 2007).  We require some evidence of X-ray emission from our
targets based on the $Rosat$ all-sky survey, and we are able to
confirm the presence of extended X-ray emission and to measure its
parameters after dedicated \emph{XMM-Newton} observations of the
selected candidates.  In this paper, we present the analysis of
combined optical and X-ray data, including measurements of
temperature, metallicity, and X-ray luminosity for three X-ray bright
groups with giant elliptical galaxies in the center.  Whenever our
analysis requires taking into account cosmological parameters, we
assume $H_0=71$ km/s Mpc$^{-1}$, $\Omega_M=0.3$, and
$\Omega_\Lambda=0.7$.

\section{Data}

In this section we describe the algorithm we used to identify X-ray
bright groups with a central giant elliptical galaxy in the SDSS data
base, and we present the \emph{XMM-Newton} data reduction process.

\subsection{Selection Criteria}

Fossil-like galaxy groups with a central giant elliptical galaxy were
most often identified based on the detection of extended hot X-ray
emission followed by optical observations. The SDSS data archive
provides the opportunity to search for such systems starting from the
optical data. The wealth of data from surveys like the SDSS and
2dFGRS, not to mention future wide-field surveys, allows for a true
systematic study of properties and evolution of groups of galaxies.
Our selection algorithm focuses on bright ellipticals that are neither
completely isolated, nor in typical galaxy groups.  The algorithm
includes the following steps.

\begin{enumerate}
\item{Identify all galaxy pairs in the SDSS DR2 spectroscopic sample
    \cite{2004AJ....128..502A} where the brighter of the two galaxies
    is at least 2 magnitudes brighter than its counterpart and $M_r <
    -22.5$.}
\item{Ensure that the bright galaxy is an early-type and has a measurable
    velocity dispersion. Choose only the most massive ellipticals. We
    choose those with velocity dispersions $> 200 $km s$^{-1}$.}
\item{Ensure that there are three galaxies within 1 Mpc, but that the
    tenth nearest neighbor is $>$1.5 Mpc away (this is $\sim$ one
    Abell radius).  This final criteria ensures that the fossil group
    candidate resides in a slightly overdense region, but not in a
    generic group or cluster.}
\end{enumerate}

The above selection criteria have consequences. First, the use of the
SDSS spectroscopic sample allows us to search a complete (at the 90\%
level) magnitude limited galaxy survey.  However, the magnitude limits
and the $\delta m_{12} =2$ requirement result in a redshift limit of
$z \sim 0.09$ for fossil group candidates. Second, the requirement of
at least 3 galaxies with 1~Mpc of the bright early-type galaxy implies
that we will not find any Isolated Over-luminous Elliptical Galaxies
(IOLEGS -- as defined by Yoshioka et al.~2004).

Applying these criteria we obtained a list of 14 candidates.  We
visually examined all of these candidates and removed two systems that
had been affected by fiber collisions in the SDSS (which only observed
85\% of galaxy pairs closer than 55 arcseconds). We matched the
remaining 12 candidates with the available X-ray information using
HEASARC and found seven RASS Faint Source Catalog matches. One of
these had been already observed (UGC00842, also known as
MS~$0116.3-0115$).

Our selection criteria were not designed to create a statistical
sample of fossil groups. Our goal was to identify massive ellipticals
in slightly over-dense regions that lack bright nearest neighbors but
are comparatively bright in X-rays.  We then targeted these
ellipticals to collect X-ray data and study their derived spectral
properties (e.g., temperatures and masses) with respect to their
optical properties.

Our optical target selection was performed in 2005.  We obtained
\emph{XMM} data for the two X-ray brightest objects from the selected
systems RX~J$002937.0-001218$ and RX~J$150548.7+030849$ in 2006 during
the XMM AO5 Cycle, and used archived \emph{XMM} data for UGC00842
\cite{2003xmm..prop..177M}.  Everywhere below we refer to these
objects as rxj0029, rxj1505, and ugc00842.

\subsection{XMM data reduction}

\begin{table}
  \def\d{\phantom{1}}
  \caption{Resulting exposures and scaling factors}\label{tab:exp_scale}
  \centering
  \medskip\def\arraystretch{1.15}
  \begin{tabular}{lccccccc}
    \hline
    \hline
    & \multicolumn{1}{c}{$T_{pn}$}\tablenotemark{a} & $\delta_{pn}$\tablenotemark{b} &
    $T_{MOS1}$ & $\delta_{MOS1}$ & $T_{MOS2}$ & $\delta_{MOS2}$ \\
    \hline
    rxj0029 & 6.7 & 3.68 & 15.4 & 1.72 & 15.4 & 1.71 \\
    rxj1505 & ...& ... & 18.6 & 1.63 & 17.6 & 1.63 \\
    ugc00842 & 5.2 & 1.21 & 6.3 & 0.99 & 6.1 & 0.94 \\
    \hline
  \end{tabular}
  \tablenotetext{a}{$T_{xx}$ exposure time for the \emph{xx} camera in
    kiloseconds.}
  \tablenotetext{b}{$\delta_{xx}$ ratio of observed fluxes in $10-15$ keV band
    outside FOV to that in ``blank'' sky data set for the \emph{xx}
    camera.}
\end{table}

We analyzed \emph{XMM} data from the EPIC/MOS and pn detectors.  The
observations were done in Full Frame mode using the THIN optical
filter.  Calibrated event files were generated using tasks
\emph{emchain} and \emph{epchain} from \emph{XMM-Newton} SAS V 7.0 and
the calibration database as available in July 2007 was used.

\begin{figure*}[t]
\centerline{
  \includegraphics[width=0.33\linewidth]{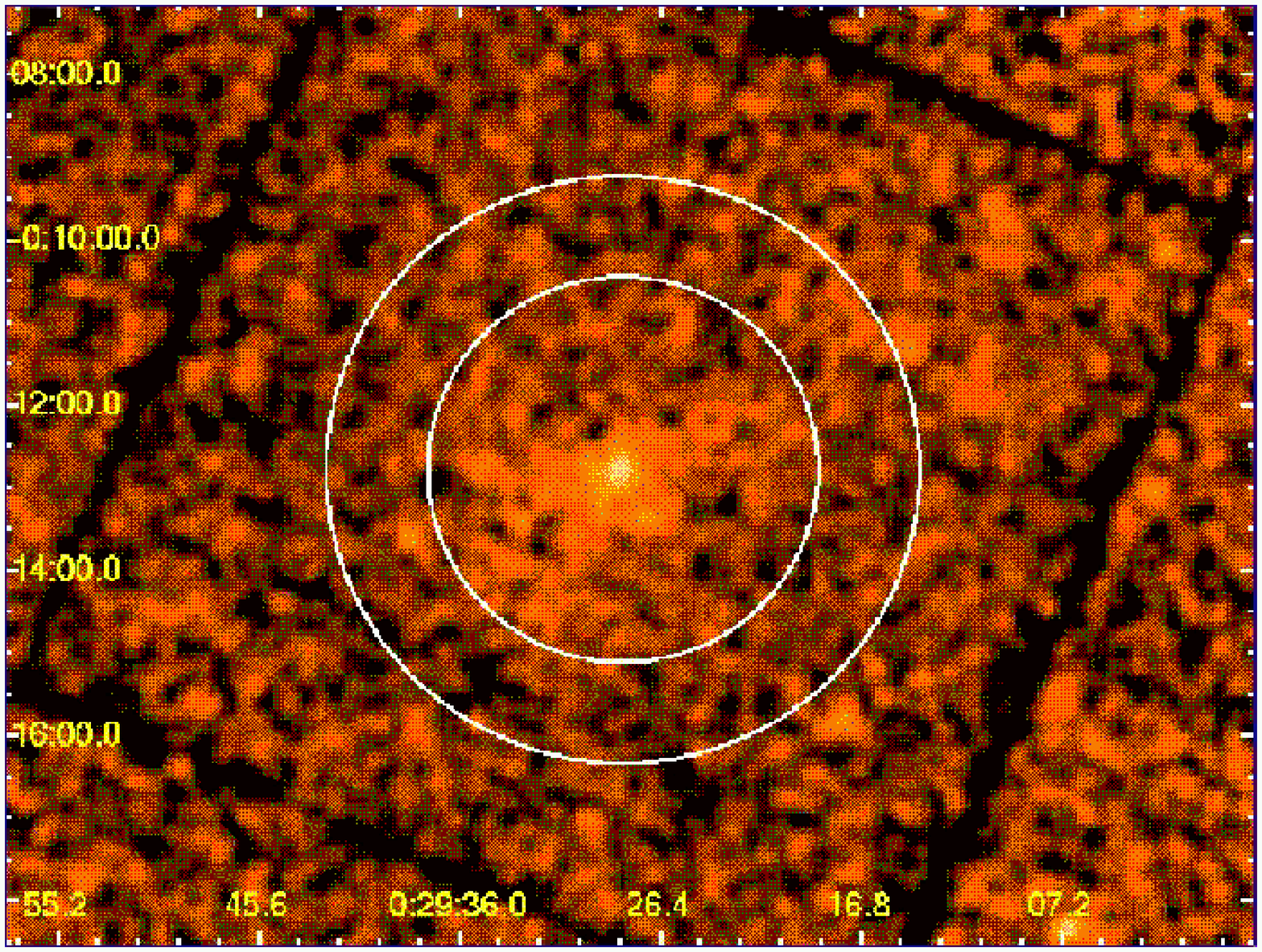}
  \includegraphics[width=0.33\linewidth]{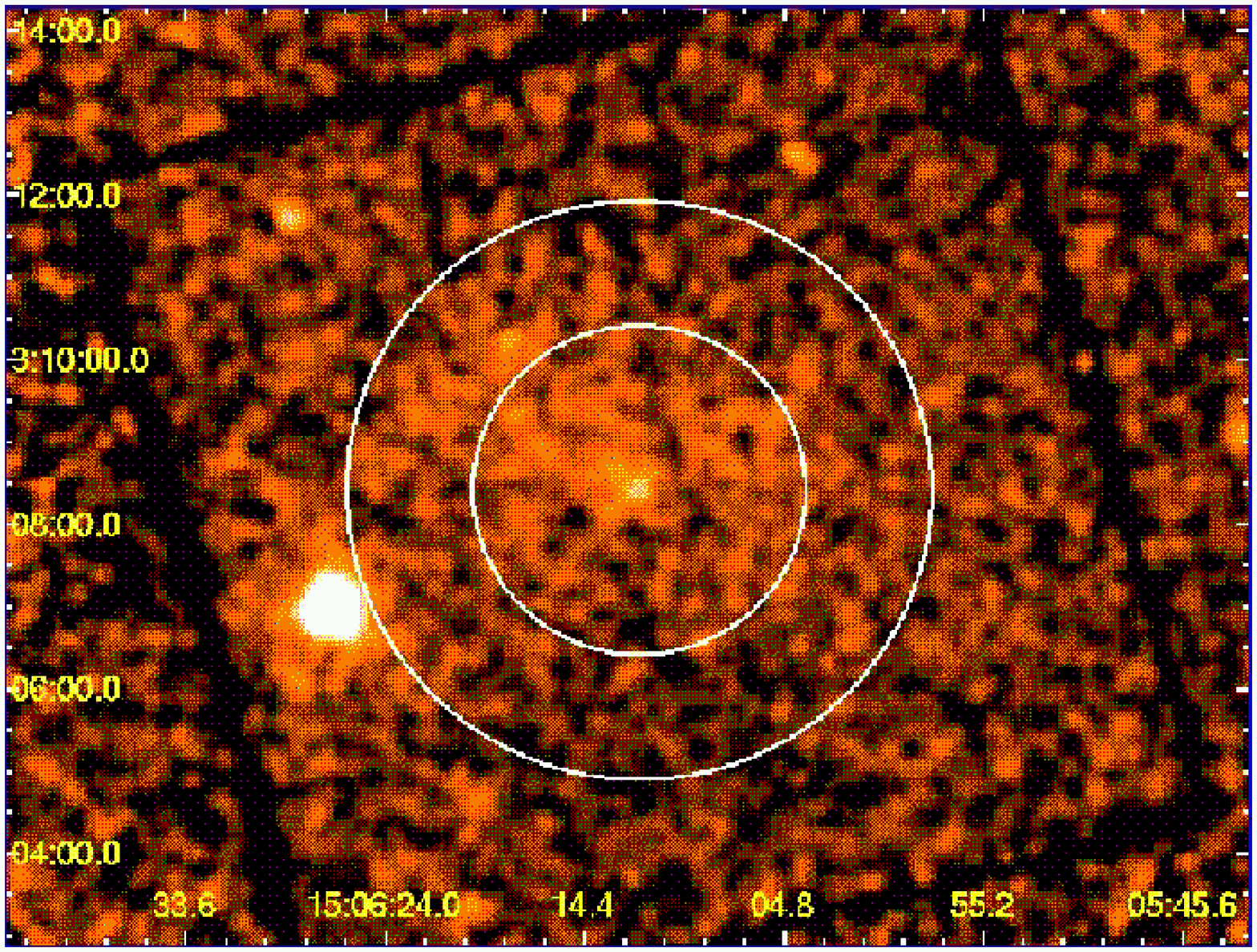}
  \includegraphics[width=0.33\linewidth]{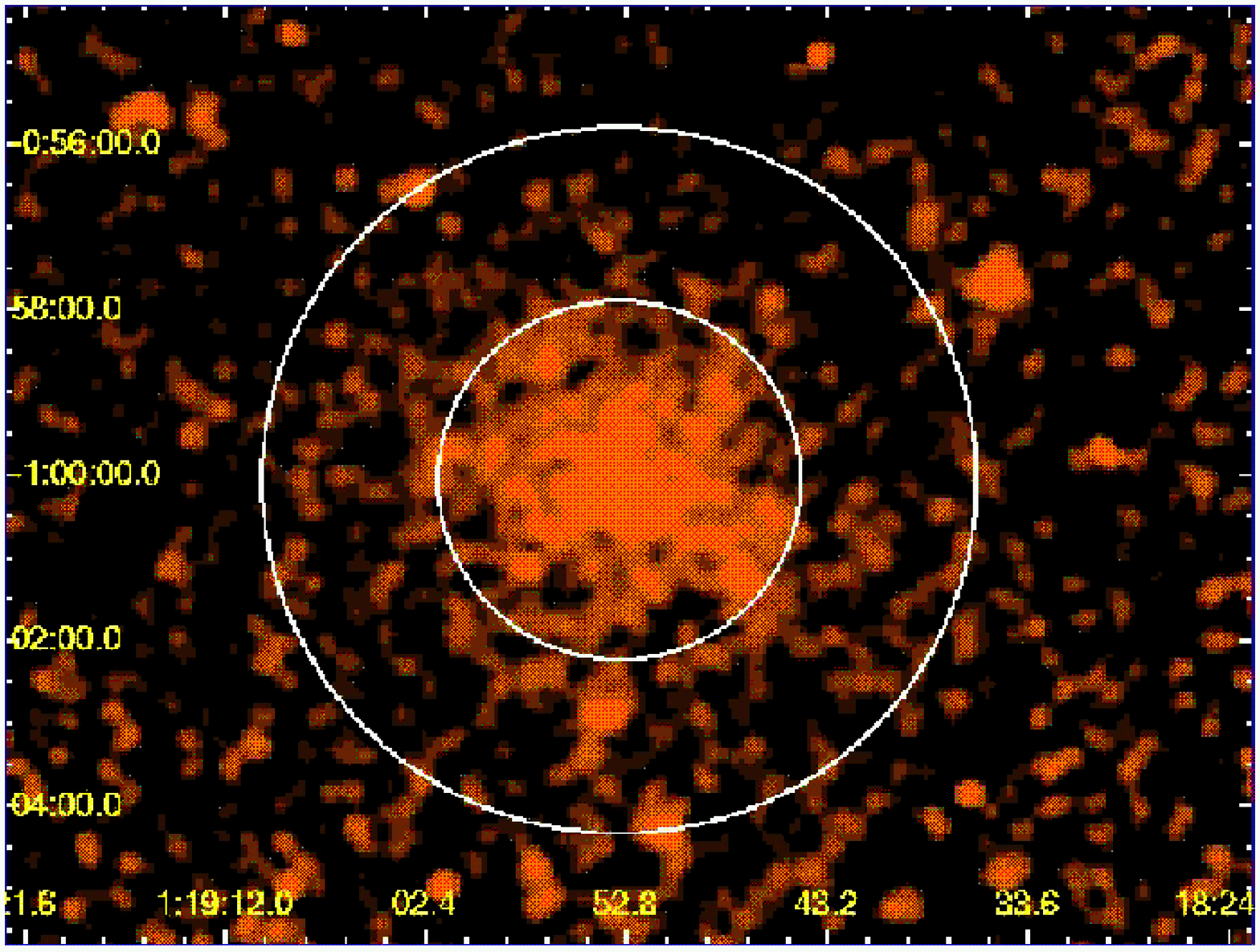}
}
\caption{X-ray images of rxj0029 (left), rxj1505 (middle), and
  ugc00842 (right) made in the $0.5-2.0$ keV band. The smoothing scale
  is 6". The inner circles, with radii 140'', 120'', and 130'', show
  the areas where the X-ray spectroscopy was performed. The outer
  circles correspond to $r_{2500}$. The bright source on the rxj1505
  image is a quasar at $z=0.21$.}
  \label{fig:images}
\end{figure*}

\begin{figure*}
\centerline{
  \includegraphics[width=0.33\linewidth]{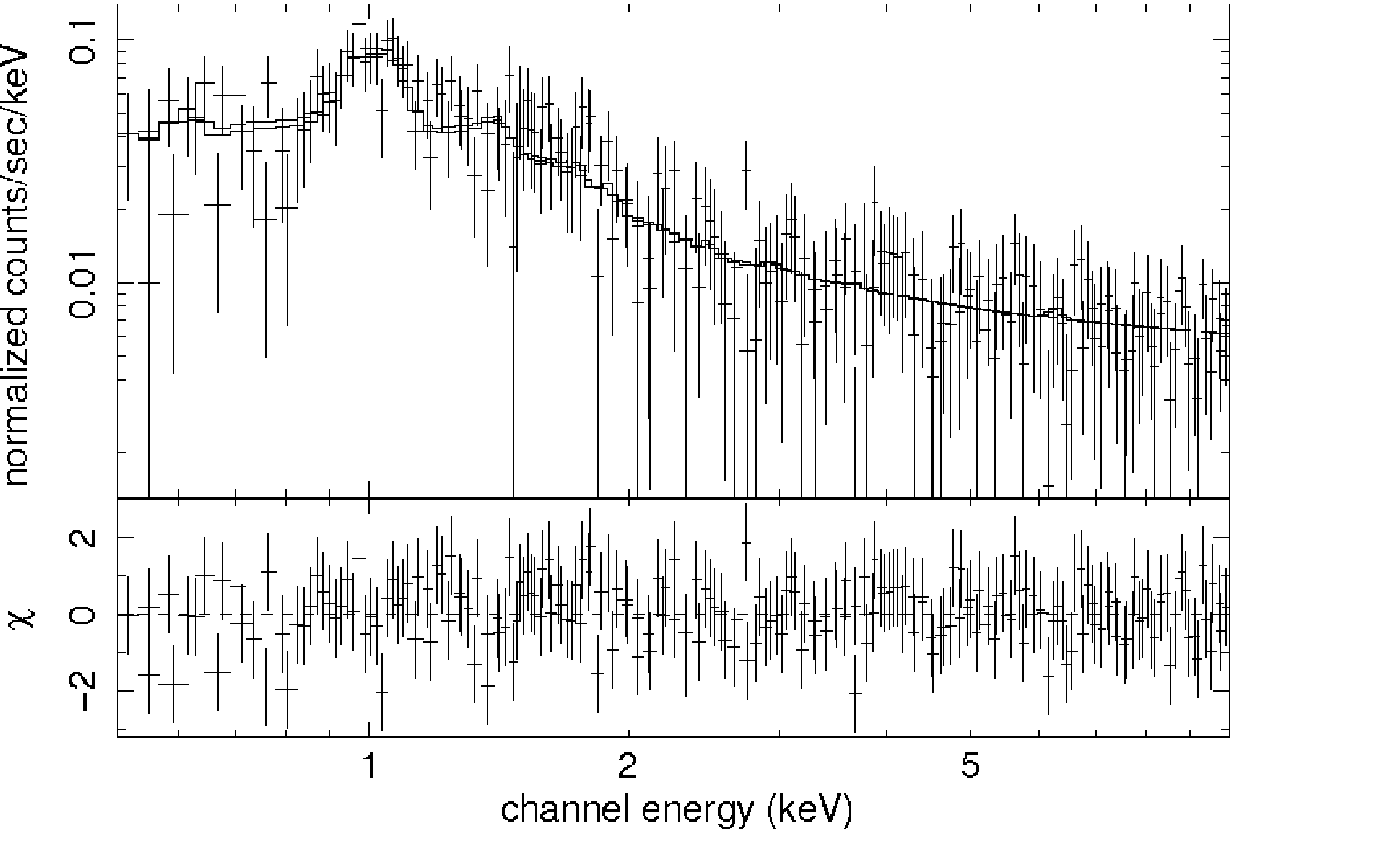}
  \includegraphics[width=0.33\linewidth]{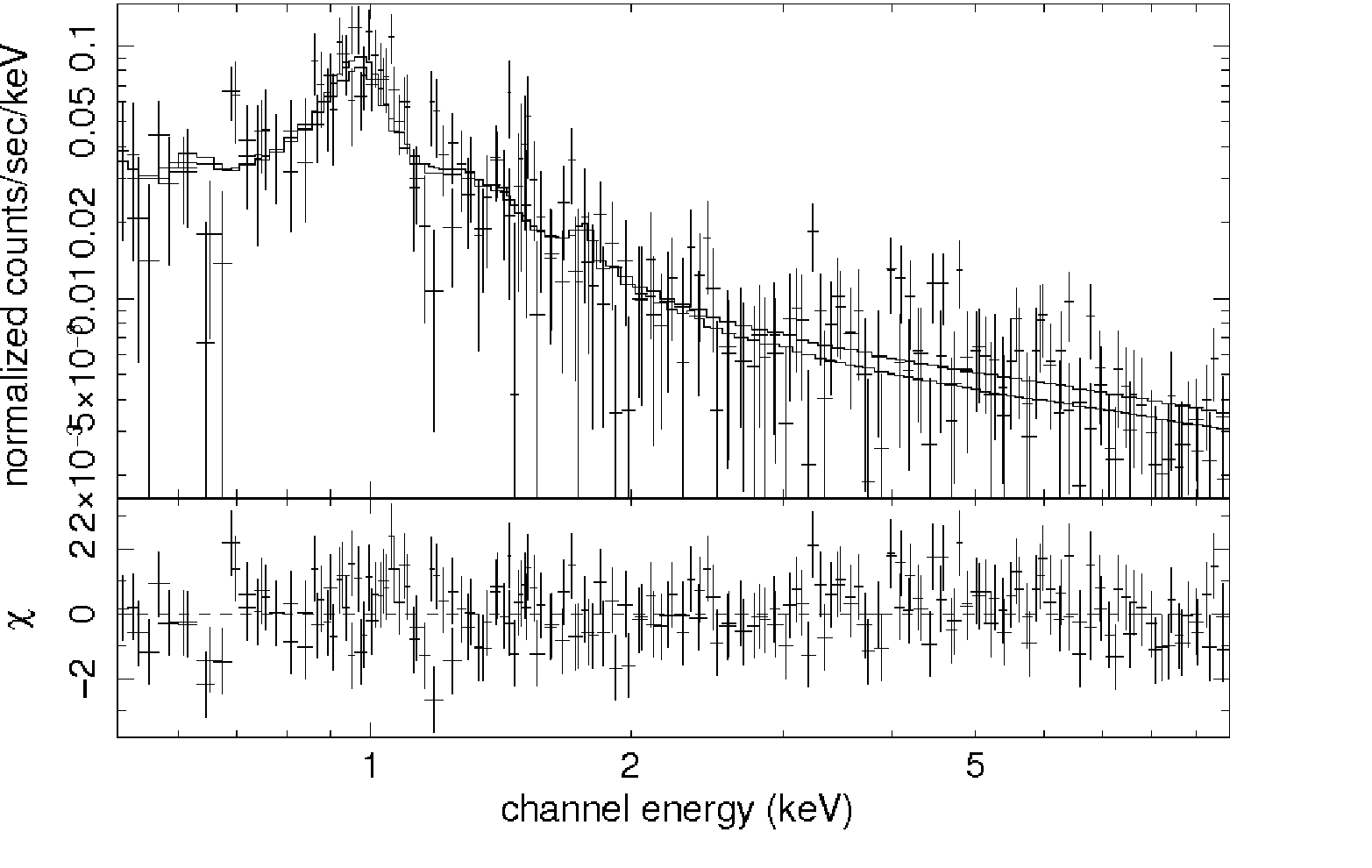}
  \includegraphics[width=0.33\linewidth]{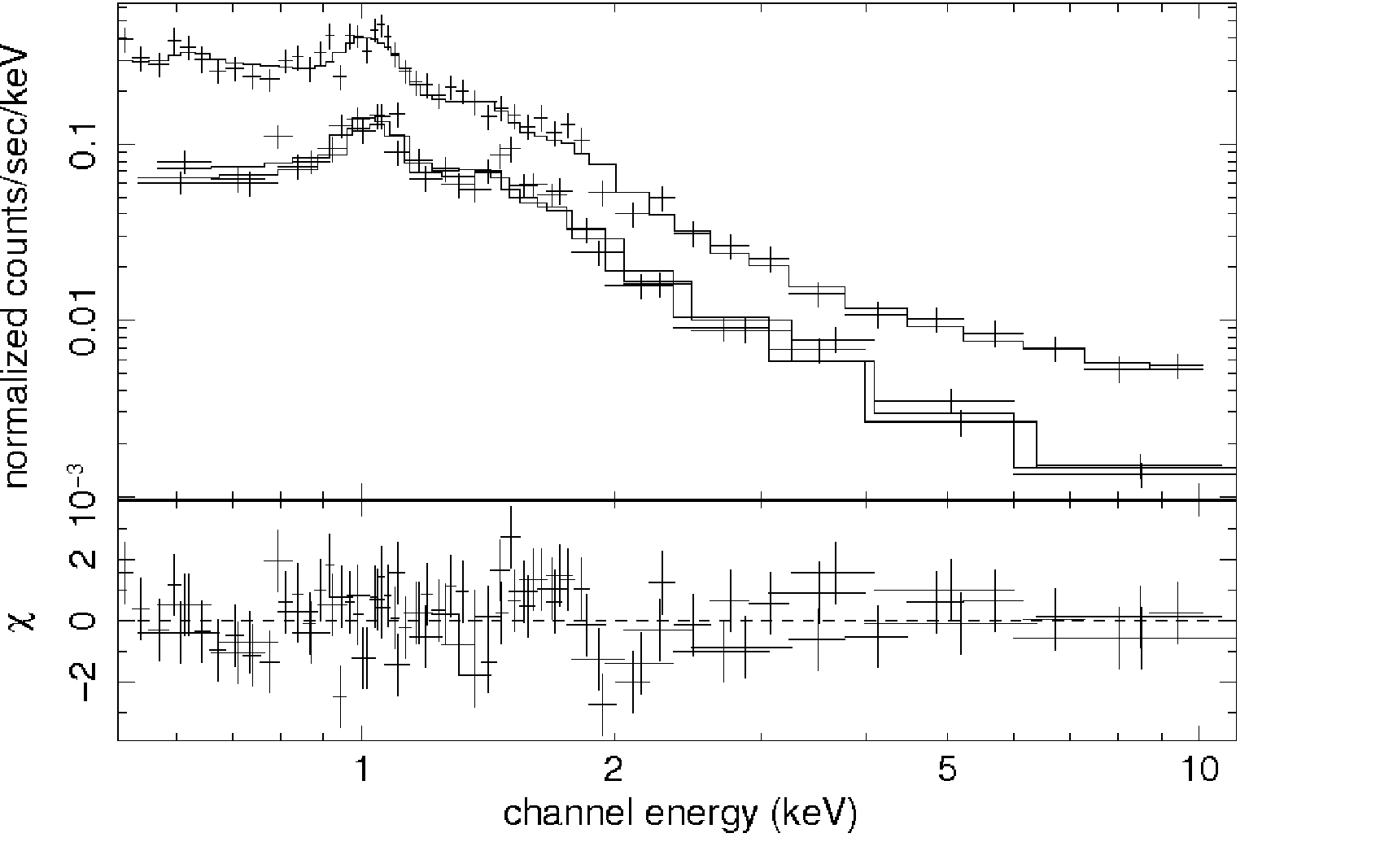}
}
 \caption{Spectra of rxj0029 (left), rxj1505 (middle) and ugc00842 (right).}
 \label{fig:spectra}
\end{figure*}

For our analysis we follow closely the data preparation technique
described in~Kotov \& Vikhlinin (2005). We summarize here the most
important steps. We use the events corresponding to patterns 0--12 for
the MOS cameras and 0--4 for the pn camera. We exclude chip \#2 in the
MOS1 data of the rxj1505 observation since it looks brighter than the
others on the image compiled from the preprocessed event
file\footnote{See~also~\emph{XMM-Newton} helpdesk:
  http://xmm.esac.esa.int/xmmhelp/EPICMOS?id=16999}. The data for each
camera is cleaned from flares using the 2--15 keV band and data from
the entire field of view is used, with the exception of point sources.
Good time intervals are generated from the periods on the lightcurves
for which deviations from the mean rate are less than $2\sigma$.
Cleaned exposure times for each camera are given in
Table~\ref{tab:exp_scale}.  The rxj1505 data from the pn camera are
highly contaminated by flares, so we do not use them.

Since the observed objects are not bright, the X-ray images contain
some regions which are dominated by background emission. Therefore, a
double-subtraction technique~\cite{2002AA...390...27A} for modeling
the background is used.  The first step of this method is a
subtraction of the energetic particles induced background. In order to
do this, normalization coefficients between the background present in
the observation and background characteristics for the given camera
are found.

We calculate these coefficients as a ratio of fluxes from the outside
field of view in the 10--15 keV band between the observation and the
``template'' background file.  As a ``template'' file for each camera,
we use a compilation of ``blank'' sky observations as described in
Carter \& Read~(2007). The normalization coefficients we obtain in
this way are listed in Table~\ref{tab:exp_scale}.  For rxj0029 and
rxj1505 the scaling coefficients are higher than 1.5.  Good values
should lie within 0.8--1.2 interval. In our case the data from both
observations are contaminated by particle background. Nevertheless, as
we show in our analysis, these observations are still useful for
imaging and spectroscopic measurements. The scaling coefficient for
the pn data of rxj0029 is too high, see Table~\ref{tab:exp_scale}, so
we did not use the pn data from this observation.

\bigskip

\section{X-ray Properties}

In this section we present the spectral and imaging analysis of our
groups.  Fig.~\ref{fig:images} shows the smoothed, particle background
subtracted, vignetting and exposure corrected images of rxj0029,
rxj1505, and ugc00842.

\subsection{X-ray Spectral Analysis}

\begin{table*}
\begin{center}
  \def\d{\phantom{1}}
  \caption{Spectroscopic parameters}\label{tab:spec_par}
  \centering
  \medskip\def\arraystretch{1.15}
  \begin{tabular}{lcccccc}
    \hline
    \hline
    & \multicolumn{1}{c}{$z$} & $F_X$\tablenotemark{a} & $L_X$\tablenotemark{b} & $T$
    & $Z$ & $L_{X,bol}$\tablenotemark{c}\\
    Name & & ($10^{-13}$ erg s$^{-1}$ cm$^{-2}$) & ($10^{42}$ erg
    s$^{-1}$) &(keV)&  ($Z_\odot$) & ($10^{43}$ erg s$^{-1}$)\\
    \hline
    rxj0029 & 0.060& $2.83\pm0.88$& $2.41\pm0.68$& $2.10\pm0.31$ & $0.66\pm0.28$ & $1.16\pm0.17$\\
    rxj1505 & 0.042& $1.92\pm0.47$& $0.80\pm0.21$& $1.13\pm0.15$ & $0.25\pm0.10$ & $0.44\pm0.04$\\
    ugc00842& 0.045& $5.28\pm0.62$& $2.51\pm0.25$& $1.90\pm0.30$ & $0.34\pm0.12$ & $1.63\pm0.05$\\
    \hline
  \end{tabular}

  \tablenotetext{a}{X-ray flux measured in the 0.5--2.0 keV band.}
  \tablenotetext{b}{X-ray luminosity in the 0.5--2.0 keV band.}
  \tablenotetext{c}{Bolometric luminosity (0.1--20.0 keV band) obtained by
    extrapolation of the measured luminosity to $r_{500}$ using the
    $\beta$-model.}
\end{center}
\end{table*}

We extract the spectra of rxj0029, rxj1505, and ugc00842 in circles
with 140'', 120'', and 130'' radii, respectively (see the inner
circles in Fig.~\ref{fig:images}).

The response and effective area files are generated with \emph{rmfgen}
and \emph{arfgen} tasks from the SAS package. The resulting spectra
are binned in such a way that there are at least 40 photons in every
bin.  We use the absorbed MEKAL model~\cite{Legacy1995} as a fitting
model, where the Galactic absorption is fixed at a value obtained from
radio surveys~\cite{1990ARAA..28..215D}. However, due to the
contamination of the rxj0029 and rxj1505 data by the particle
background we have to add a power-law component without a correction
for the effective area.  This component allows us to describe the high
energy part of the spectrum. We also add the same component to the
ugc00842 spectrum in order to describe residual contaminations by
flares, since any further thorough cleaning of flares, such as
changing $\sigma$-clipping or experiments with different cleaning
energy bands, only reduced the exposure time without improving the
spectrum.

The data from MOS1 and MOS2, including pn in the case of ugc00842, are
fitted jointly in the 0.5--10 keV band, where the temperatures,
metalicities, and power-law slopes are held fixed, while the
normalizations for each spectral component and for each of the
detectors is kept free. The resulting spectra for the groups are shown
in Fig.~\ref{fig:spectra} and the derived spectral parameters are
given in Table~\ref{tab:spec_par}.

Figure~\ref{fig:spectra} shows that the main feature of the group
spectra in the soft band is a bremsstrahlung component. The power-law
component in this band is several times lower. The values for
temperatures and abundances are primarily determined from the shapes
of the spectra in this band. Hence, the parameters inferred from the
fits are quite reliable even for the groups rxj0029 and rxj1505 which
are contaminated by particle induced backgrounds.

\begin{figure*}
\centerline{
  \includegraphics[width=0.33\linewidth]{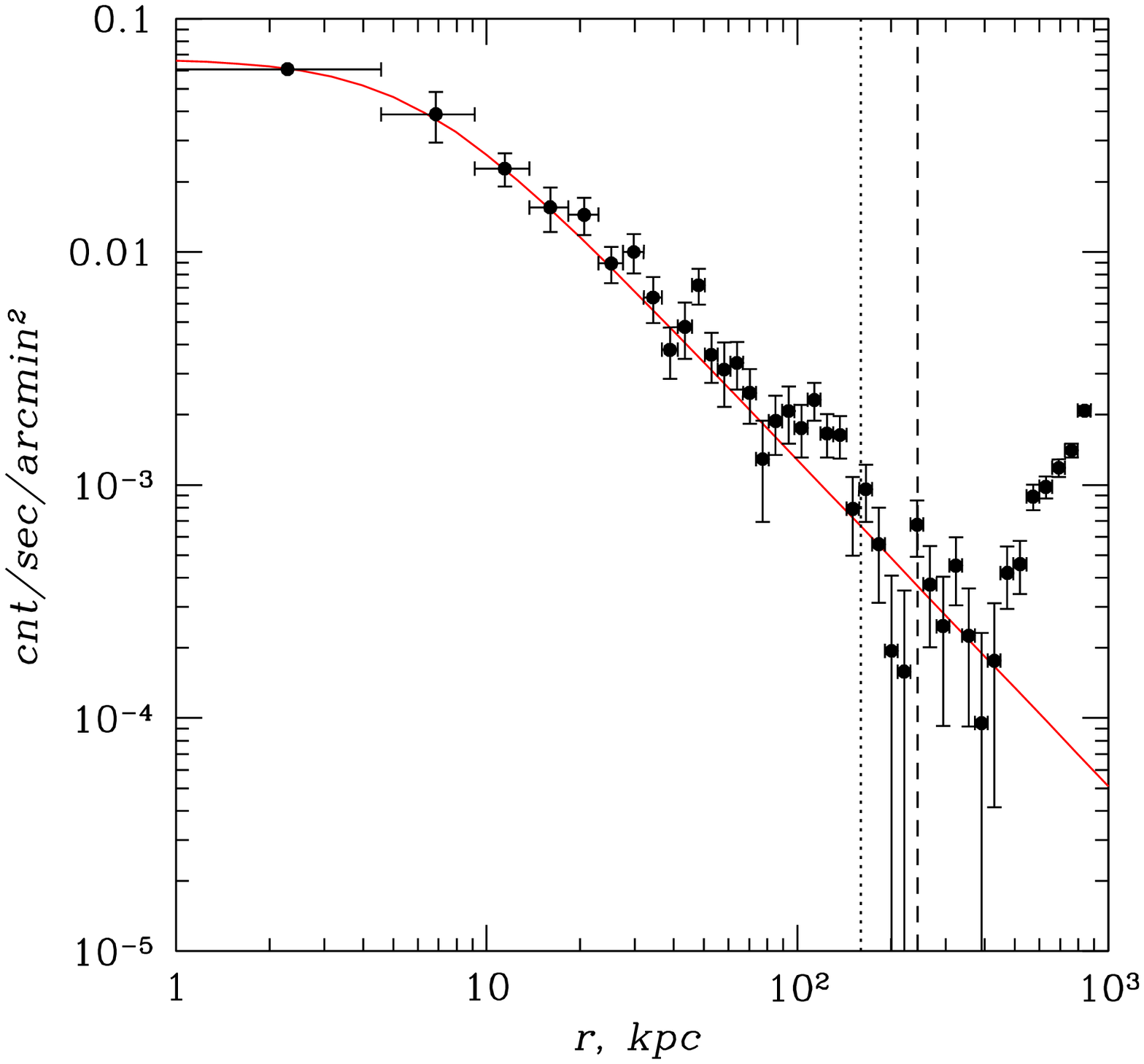}
  \includegraphics[width=0.33\linewidth]{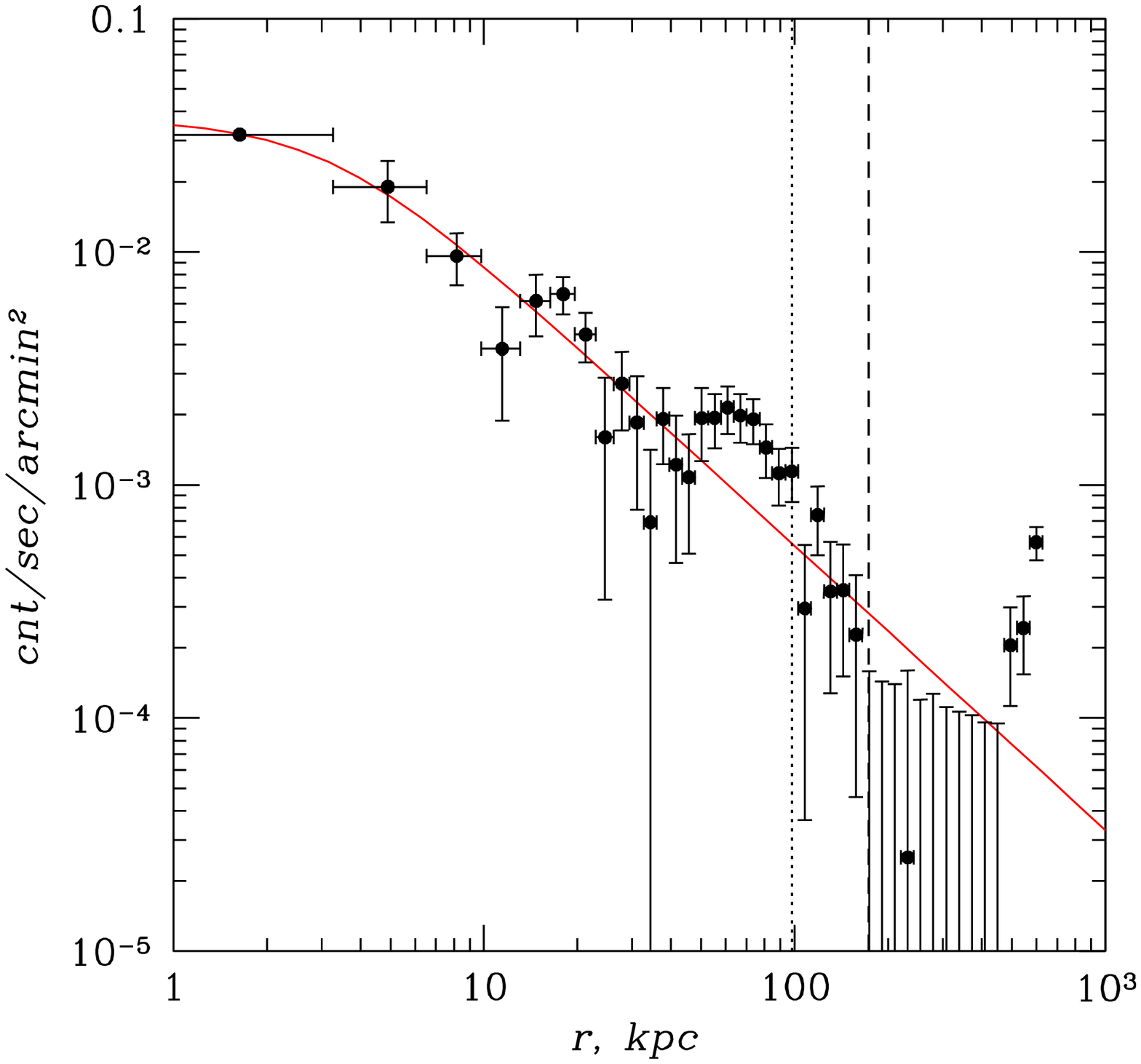}
  \includegraphics[width=0.33\linewidth]{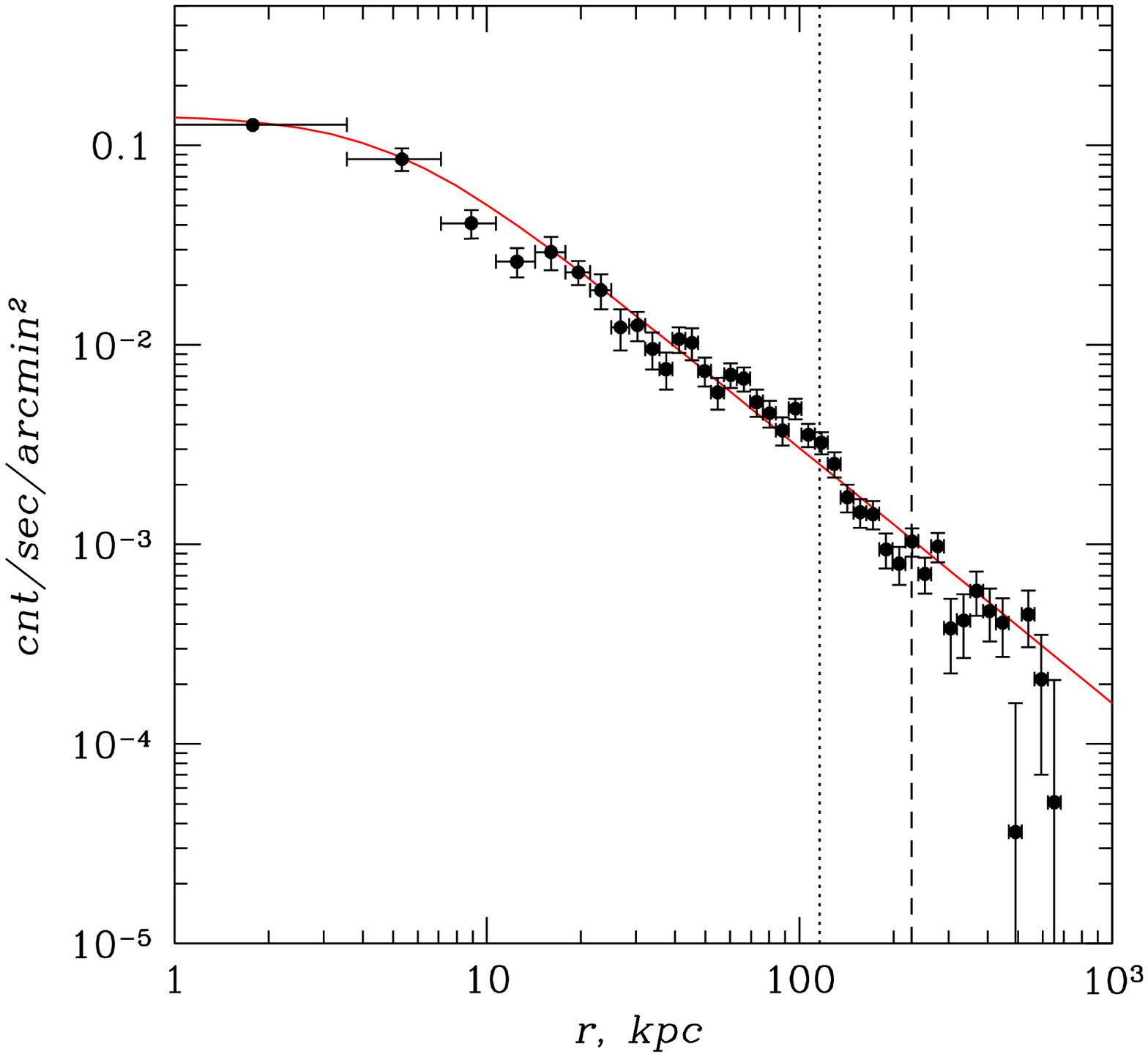}
}
\caption{Surface brightness profiles of rxj0029 (left), rxj1505
  (middle), and ugc00843 (right). The dotted lines show the regions
  where temperatures and other spectroscopic parameters were measured.
  The dashed lines show $r_{2500}$.}
  \label{fig:profiles}
\end{figure*}

\subsection{X-ray Imaging Analysis}
For the imaging analysis in the 0.5--2.0 keV energy band we mask all
detectable point sources on the MOS1 and MOS2 detectors for three of
our groups. Then we add vignetting corrected images. Surface
brightness profiles are shown in Fig.~\ref{fig:profiles}. We use the
$\beta$-model ~\cite{1976AA....49..137C} to fit the data:
\begin{equation} S(r)=S_0 (1+ \frac{r^2}{{r_c}^2})^{-3\beta+0.5},
\end{equation}
where $r$ is the angular projected off-center distance and $r_c$ is
the core radius of the distribution.  The best fit parameters for
$r_c$ and $\beta$ are given in Table~\ref{tab:prof_par} and the
surface brightness profiles extracted from the complete field of view
are shown in Fig.~\ref{fig:profiles}. The derived parameters $r_c$ and
$\beta$ fall in the range of values for groups of galaxies measured in
Osmond \& Ponman 2004. Several outer points on the brightness profiles
for rxj0029 and rxj1505 lie higher than one would expect, especially
for rxj0029. This effect is probably caused by the particle induced
background and vignetting corrections. The same effects but in lesser
degree, can also be seen in Kotov \& Vikhlinin~(2005).  These last
points give some idea of the possible uncertainties in our results due
to experimental errors.  As one can see, most of our points lie well
above this level, and our fit parameters are stable against these
effects.

Assuming a spherically symmetric density distribution and hydrostatic
equilibrium of the intragroup gas, and using the fitted temperatures
and fit parameters of the $\beta$-model, we estimate the total masses
of the groups inside $r_{2500}$ and $r_{500}$ as
\begin{equation}
M(< r_{2500(500)}) = 1.1 \times 10^{14}M_\odot T_{keV} \beta
\frac{r_{2500(500)}^3}{r^2_{2500(500)}+r_c^2}
\end{equation}
The estimated values for $M_{t,2500}$ and $M_{t,500}$ (see
Table~\ref{tab:prof_par}) are typical for groups of galaxies.

Knowing $r_{2500}$ and $r_{500}$ we can obtain gas mass estimations
inside these radii.  We use the popular deprojection technique
\cite{1981ApJ...248...47F, 1997MNRAS.292..419W} in order to obtain
volume emissivity, which is then converted to the gas density and gas
mass.  Since the brightness profiles are quite noisy, we deproject the
$\beta$-model fits rather than count rates. The results are shown in
Table~\ref{tab:prof_par}.

The gas fractions inside $r_{2500}$ and $r_{500}$ are in good
agreement with the gas fractions for other groups of galaxies with the
same temperatures (see Fig.~4 and Fig.~6 in Sanderson et al.~2003).

\thispagestyle{empty}
\begin{table*}
\begin{center}
  \caption{Parameters inferred from the X-ray data}\label{tab:prof_par}
  \begin{tabular}{lcccccccccc}
    \hline
    \hline
    & \multicolumn{1}{c}{$r_c$} & $\beta$ & $r_{2500}$\footnote{$r_{2500}$($r_{500}$) is the  
      radius inside which the mean density of the object is 2500(500) times
      higher than the mean density of the Universe.} &  $M_{g,2500}$ &
    $M_{t,2500}$ & $f_{g,2500}$ & $r_{500}$  & $M_{g,500}$ &
    $M_{t,500}$ & $f_{g,500}$\\ 
    Name & (kpc) &  & (kpc) & ($10^{11}\,M_\odot$) &
    ($10^{13}\,M_\odot$) & ($M_g/M_t$) & (kpc) & ($10^{12}\,M_\odot$)
    & ($10^{13}\,M_\odot$) & ($M_g/M_t$) \\ 
    \hline
rxj0029 & $5.9\pm0.5$ &$0.40\pm0.03$ & 243 & $5.1\pm1.4$ & $2.34\pm0.40$ & ~2.2\% & 545 &
    $2.2\pm0.6$ & $5.01\pm0.85$ & ~4.4\% \\
rxj1505 & $3.2\pm0.6$ &$0.37\pm0.05$ & 173 & $1.8\pm1.1$ & $0.79\pm0.15$ & ~2.3\% & 388 &
    $0.9\pm0.5$ & $1.78\pm0.36$ & ~5.1\% \\
ugc00842& $5.0\pm0.5$ &$0.38\pm0.02$ & 228 & $7.5\pm2.3$ & $1.70\pm0.30$ & ~4.4\% & 509 &
    $3.5\pm1.1$ & $4.03\pm0.69$ & ~8.8\% \\
    \hline
  \end{tabular}
\end{center}
\end{table*}

\section{Optical Properties}

\begin{table}
  \caption{Optical Parameters}\label{tab:optical_par}
  \begin{tabular}{lccccccc}
    \hline
    \hline
    Name\tablenotemark{a} & {$N_{spec}$}& {$\sigma_V$} & {$M_{R_c}^{BG1}$} & {$M_{R_c}^{BG2}$} & $M_{R_c}^{BG3}$&
    $L_{opt,tot}$\\ 
    {} & {} & {(km/s)} & {} & {} & & {($10^{11} L_\odot$)}\\ 
    \hline
rxj0029 & 16 &434 & $-22.68$ & $-22.27$ & $-21.86$ & 3.09\\
rxj1505 & 13 &242 & $-22.55$ & $-20.86$ & $-20.79$ & 1.98\\
ugc00842& 16 &439 & $-23.01$ & $-20.02$ & $-19.99$ & 1.93\\
    \hline
 \end{tabular}

 \tablenotetext{a}{All quantities given here are calculated for galaxies inside
   $r_{500}$.}

\end{table}

The optical data are taken from the Sloan Digital Sky Survey (SDSS)
DR5 release~\cite{2007ApJS..172..634A}.  We note that the SDSS
magnitudes have been found to be underestimated for bright galaxies as
a result of the background subtraction technique utilized by the SDSS
photometric pipeline~\cite{2007ApJ...662..808L, 2007AJ....133.1741B}.
Thus, we have corrected the petrosian magnitudes using the technique
described in von der Linden et al. (2007). These corrections are
between 0.1 and 0.2 magnitudes in the $r$ and $i$-bands. We have also
applied extinction corrections and k-corrections using
$kcorrect$\footnote{http://cosmo.nyu.edu/blanton/kcorrect/} version
v4\_1 \cite{2003AJ....125.2348B}.  To compare to the literature we
transform the SDSS $r$ magnitudes to $R_{Cousins}$ using the
transformations given in Fukugita et al. (1996).

We calculate velocity dispersions for our groups using galaxies with
available spectral information (see Table~\ref{tab:optical_par}).  For
a consistency check we estimated $r_{500}$ using
\cite{1998ApJ...505...74G, 2007MNRAS.377..595K}:
\begin{equation}
r_{500} = \frac{1.2 \sigma_V}{h}.
\end{equation}
$r_{500}$ values obtained this way are within 20\% from the $r_{500}$
values we get from the X-ray profiles (Table~\ref{tab:prof_par}), and
therefore in good agreement considering the uncertainties in
both estimations.

\begin{figure*}
\centerline{
  \includegraphics[width=0.33\linewidth]{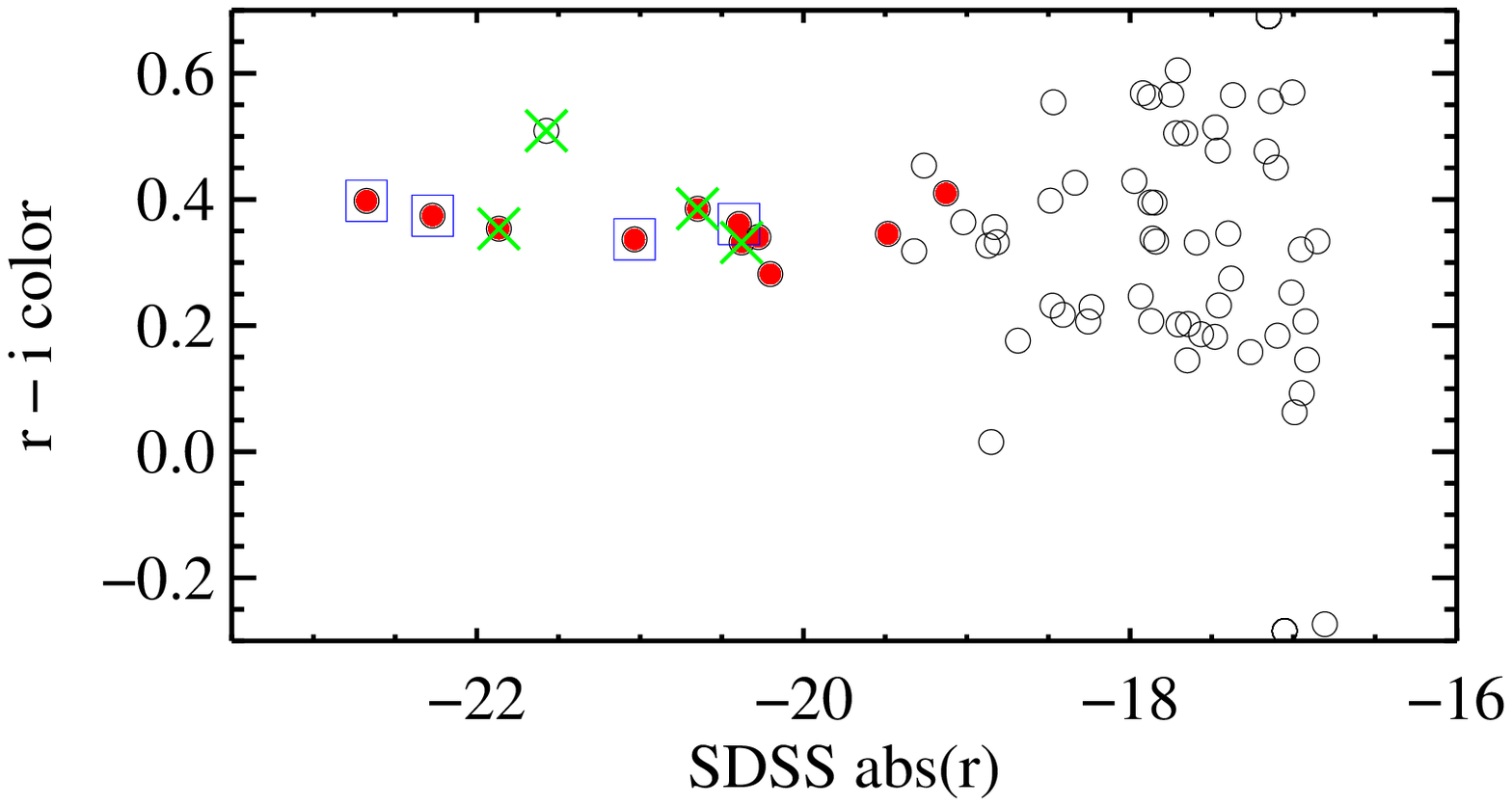}
  \includegraphics[width=0.33\linewidth]{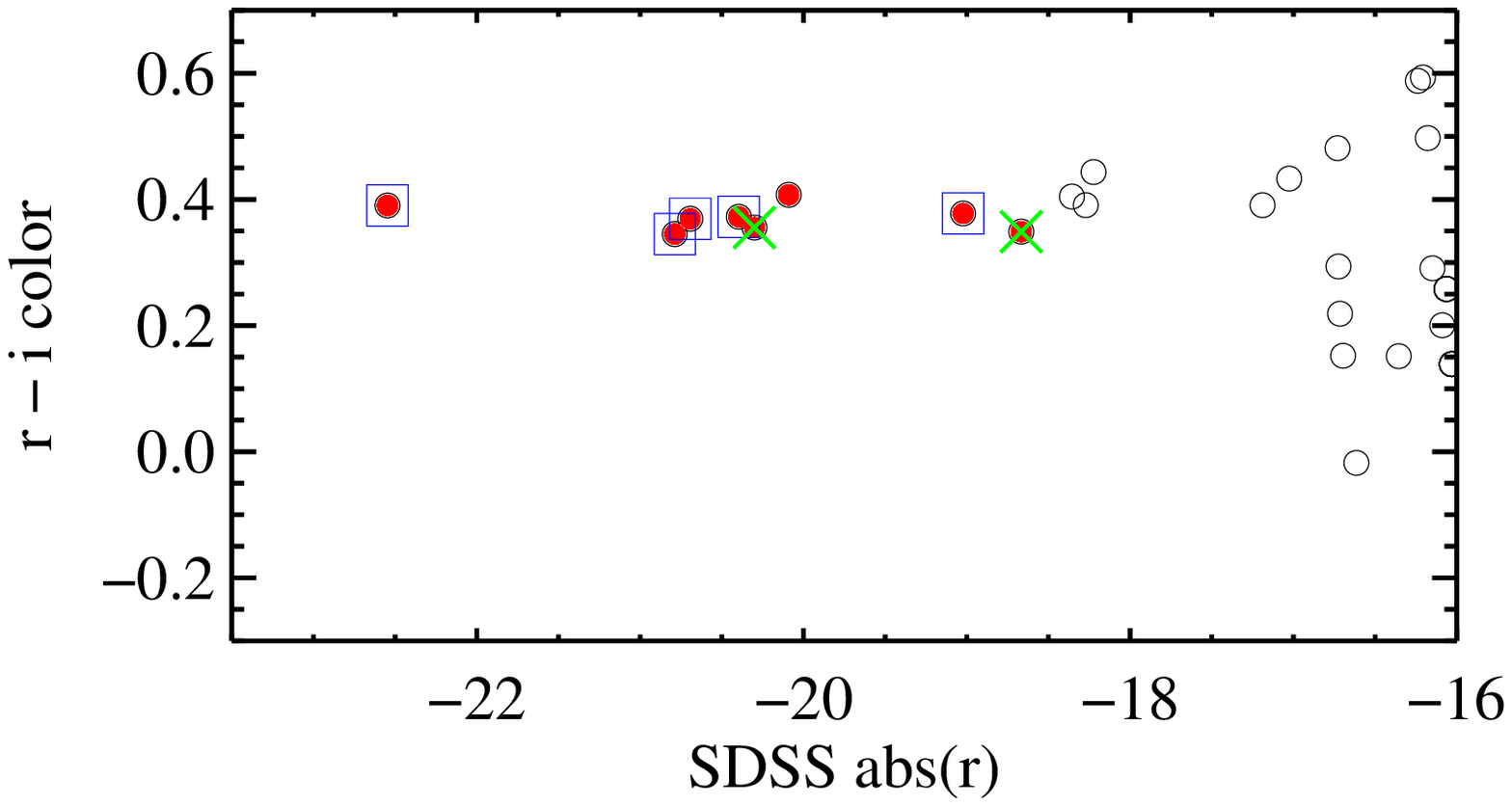}
  \includegraphics[width=0.33\linewidth]{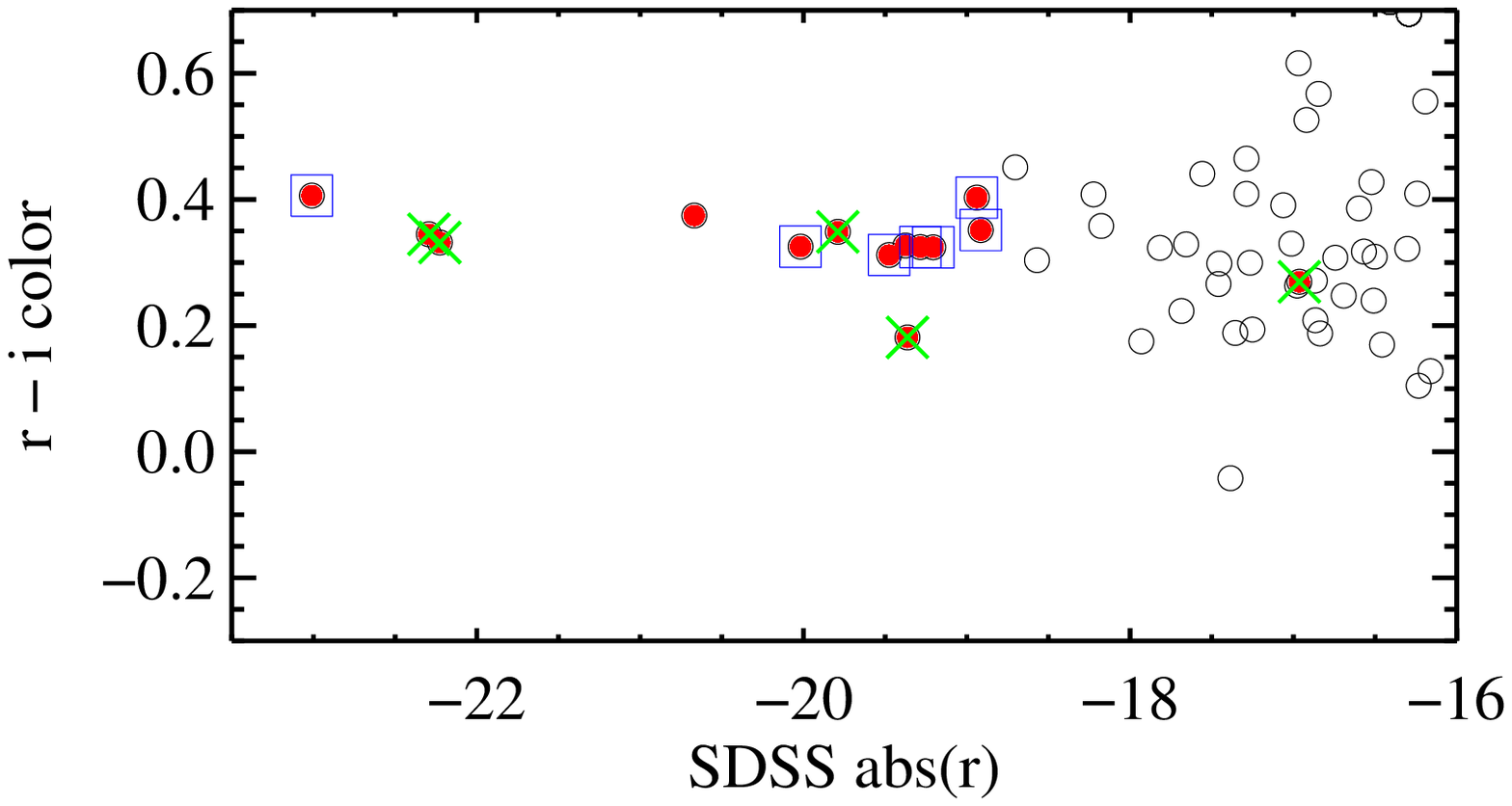}
}
\caption{The color magnitude diagram for the groups rxj0029, rxj1505,
  and ugc00842. The open circles are photometric data from the SDSS.
  The filled circles are those galaxies targeted for SDSS
  spectroscopy.  The squares show the galaxy data with observed SDSS
  velocities and that are within the velocity dispersion of the group.
  The Xs denote galaxies with observed spectra that are not group
  members.  In all figures, the region with $r_{2500}$ projected on
  the sky is used for membership.}
\label{fig:cmrs}
\end{figure*}

\section{Discussion and Conclusions}

\begin{figure*}
\centerline{
  \includegraphics[width=0.33\linewidth]{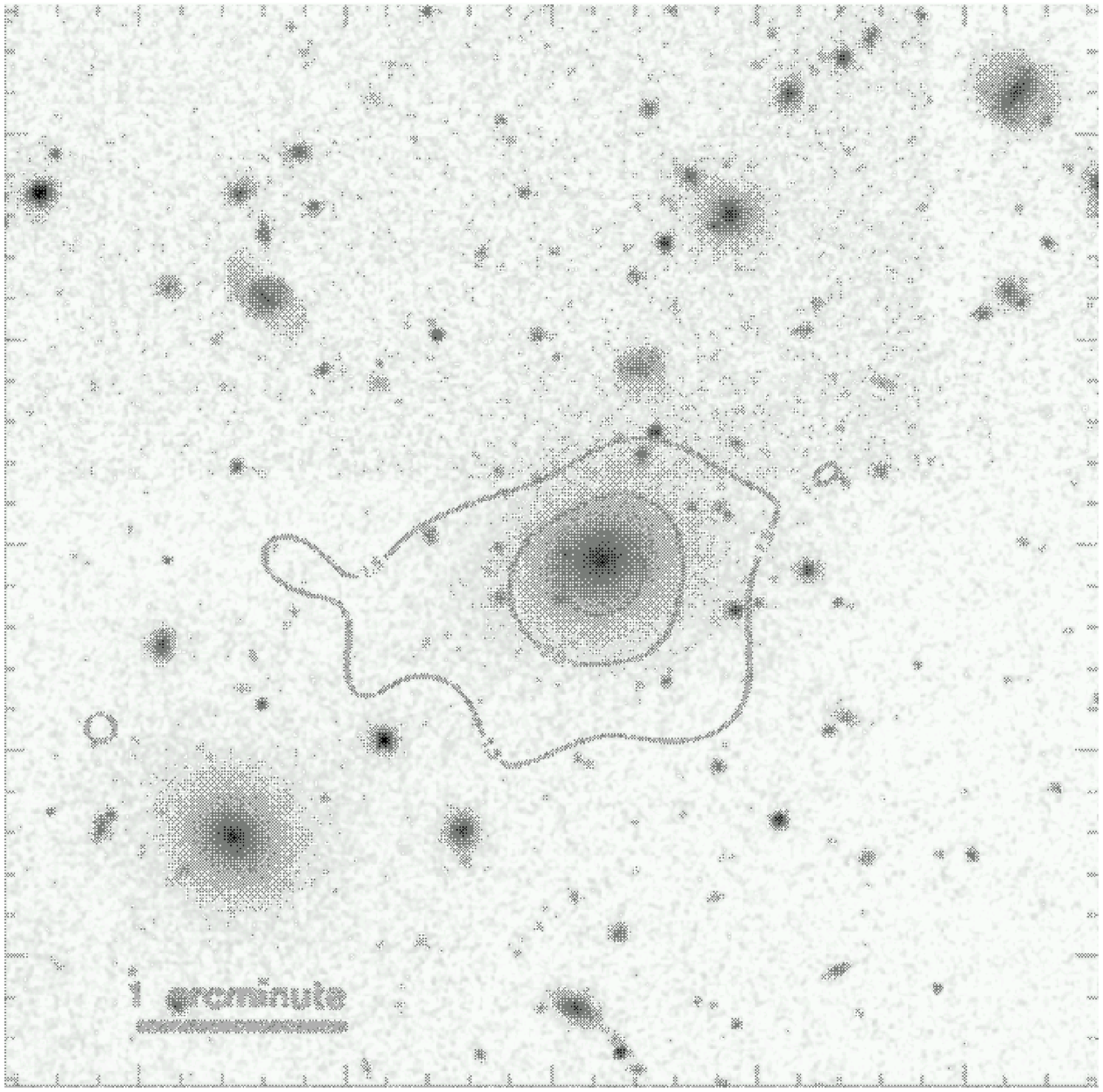}
  \includegraphics[width=0.33\linewidth]{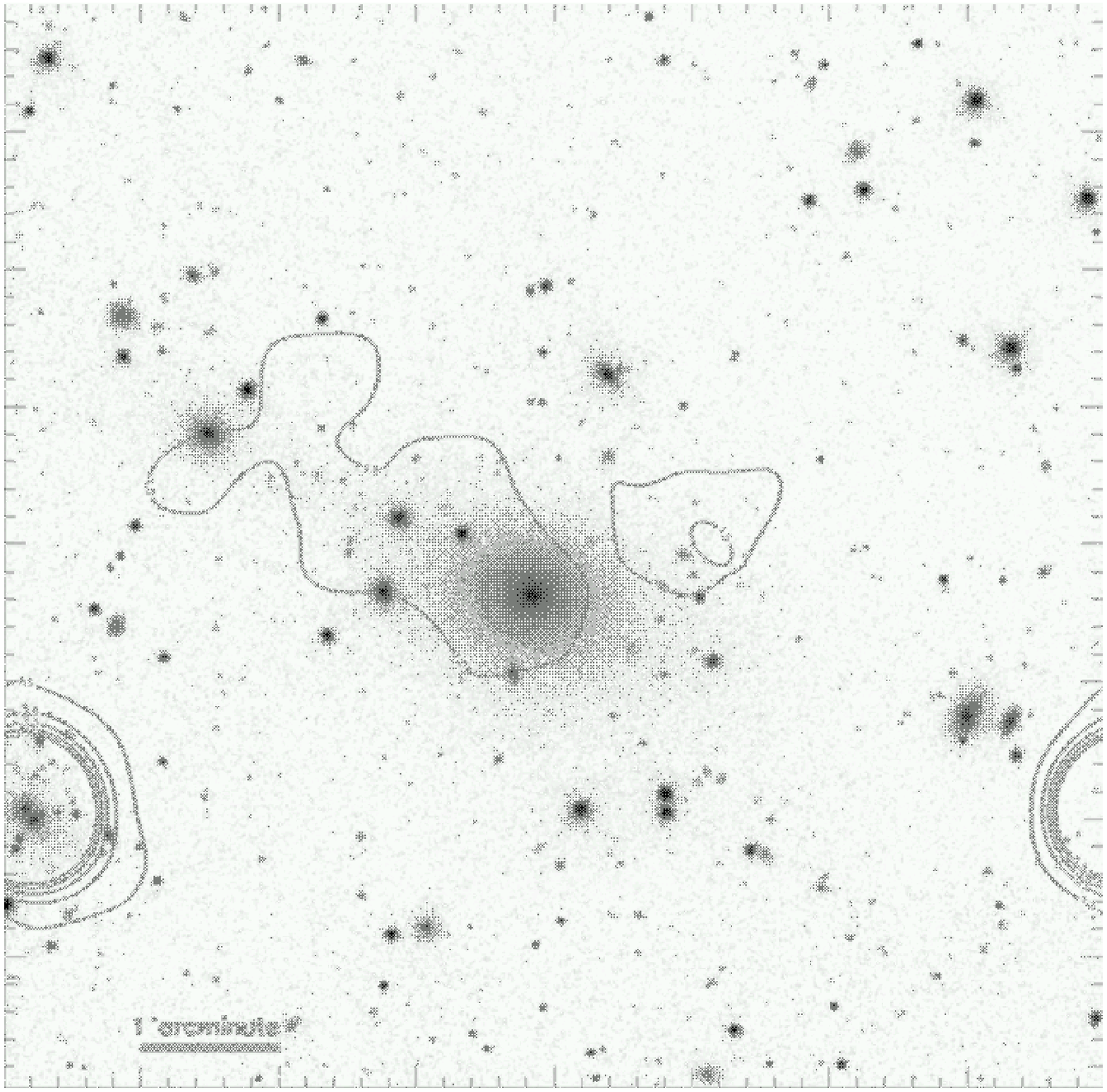}
  \includegraphics[width=0.33\linewidth]{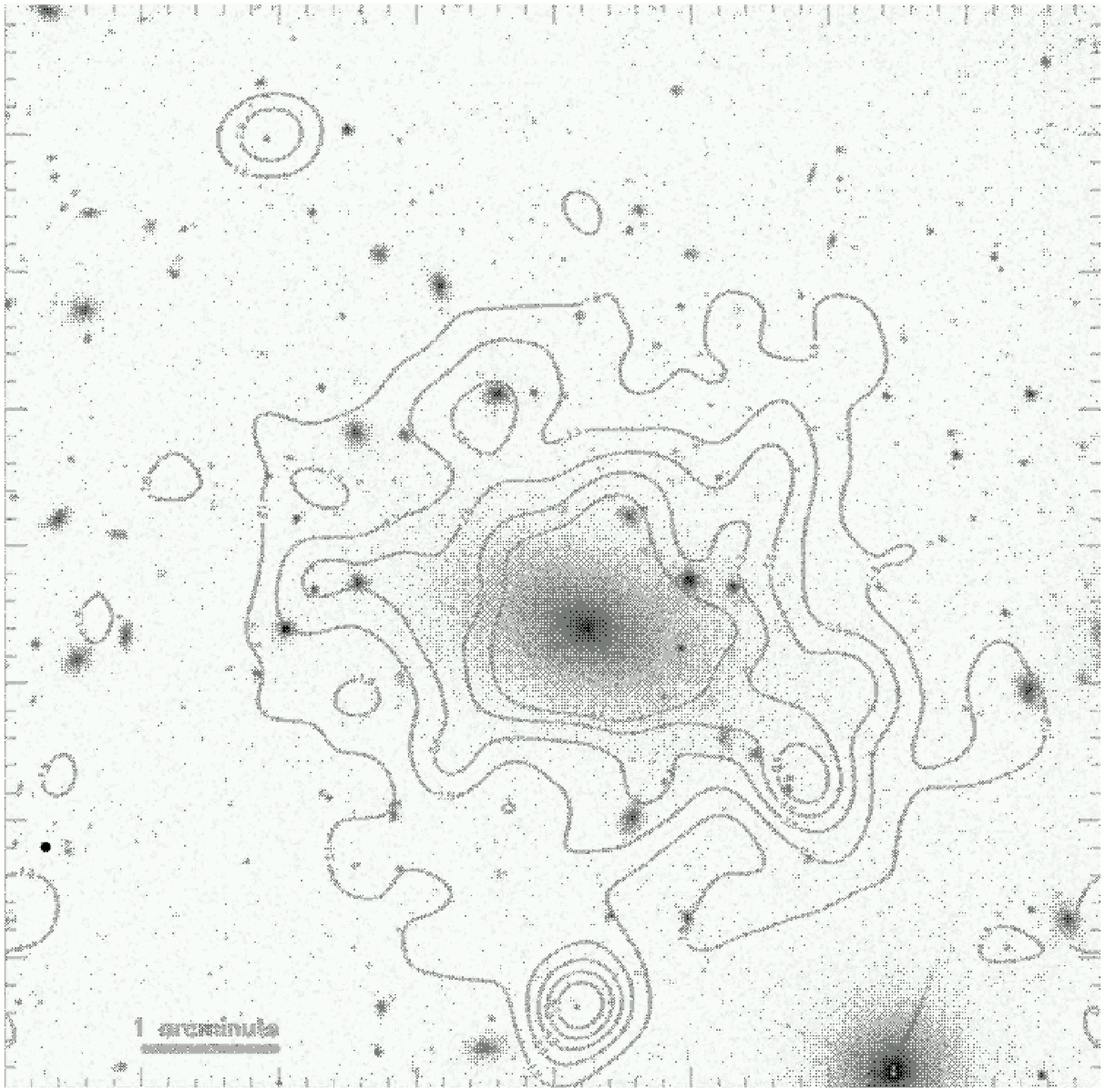}
}  
\caption{X-ray contours overlayed on the SDSS images (from left to
  right: rxj0029, rxj1505, and ugc00842).}
  \label{fig:contour_images}
\end{figure*}

\begin{figure*}
  \plottwo{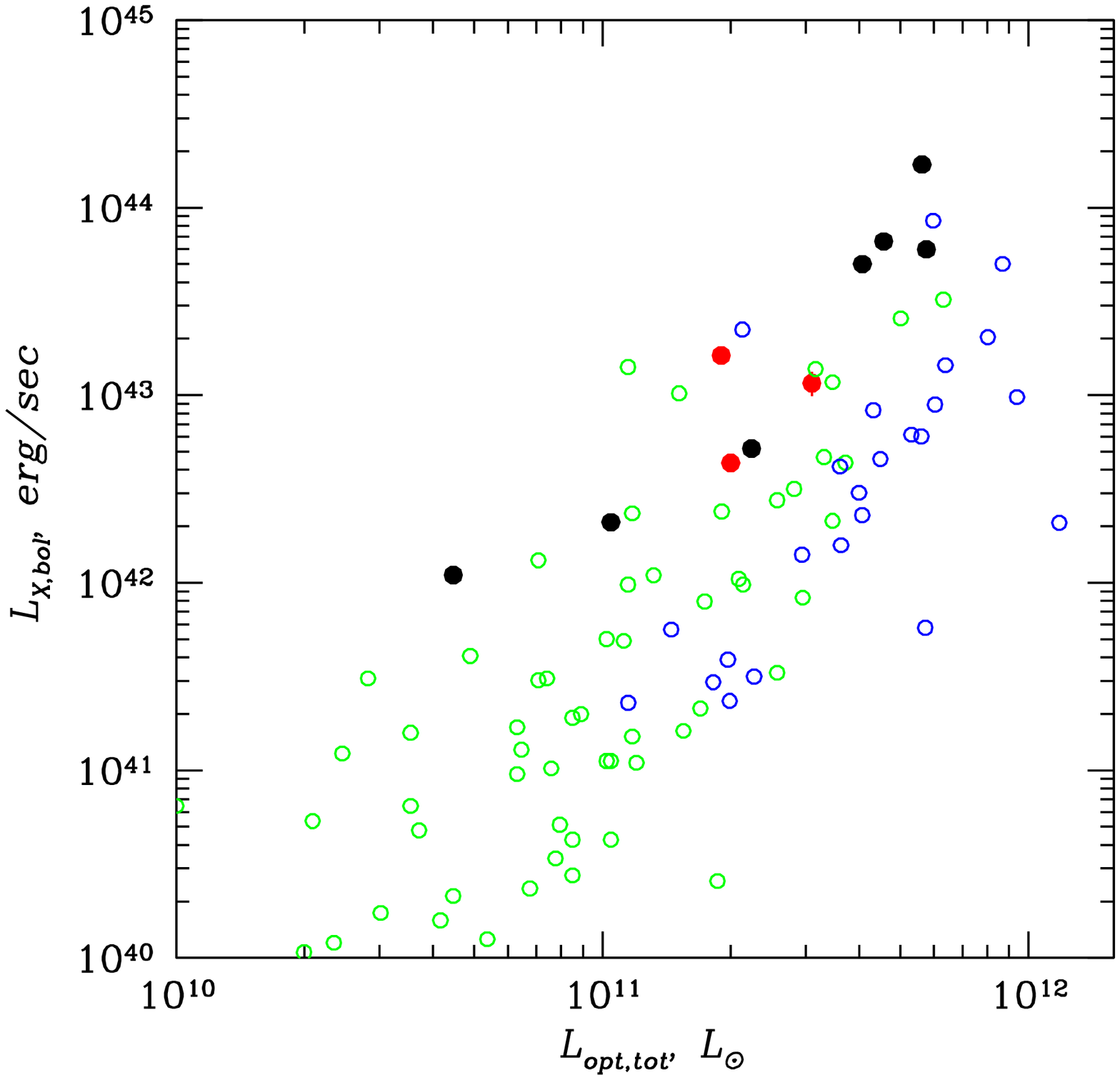}{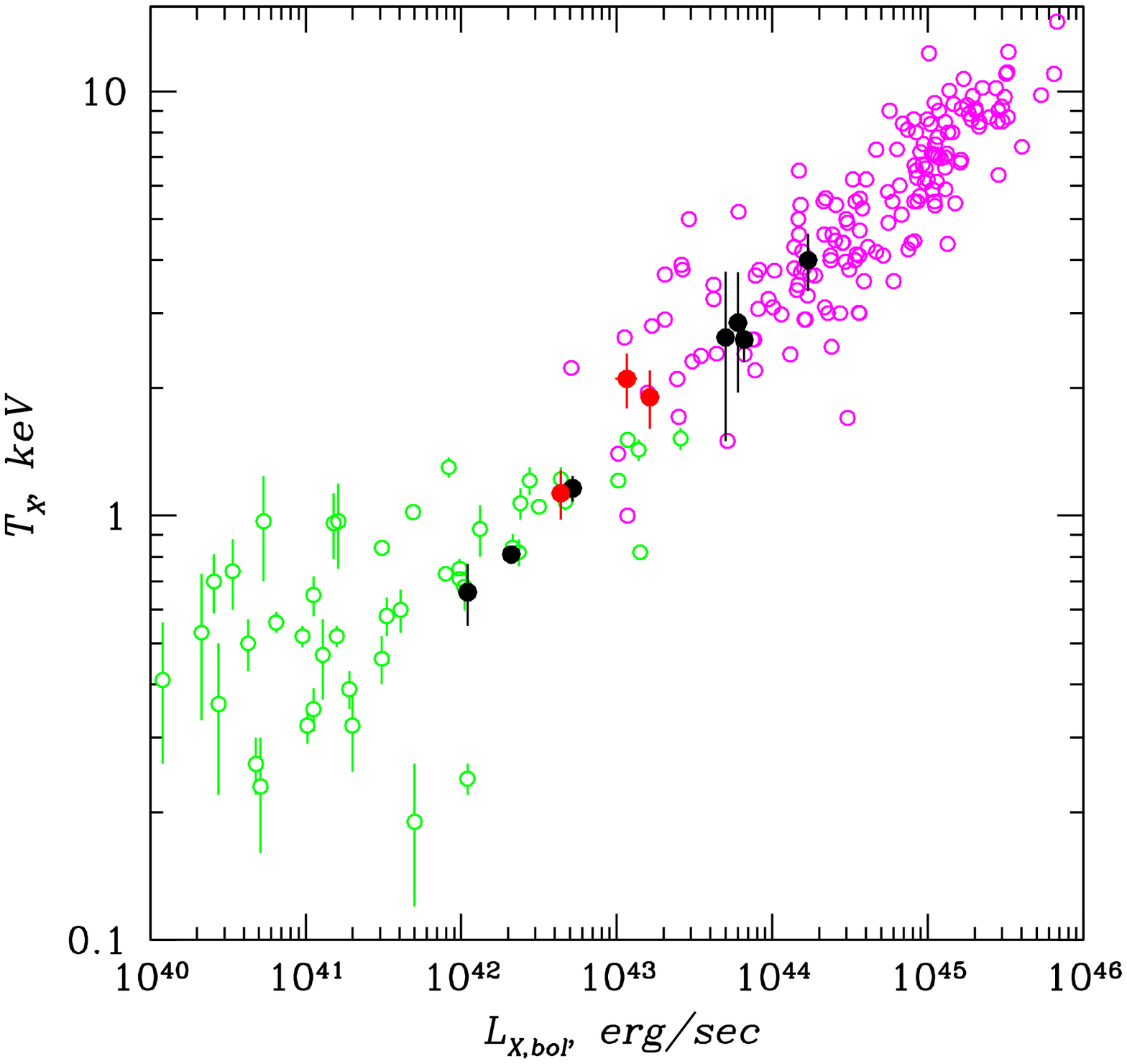}
  \plottwo{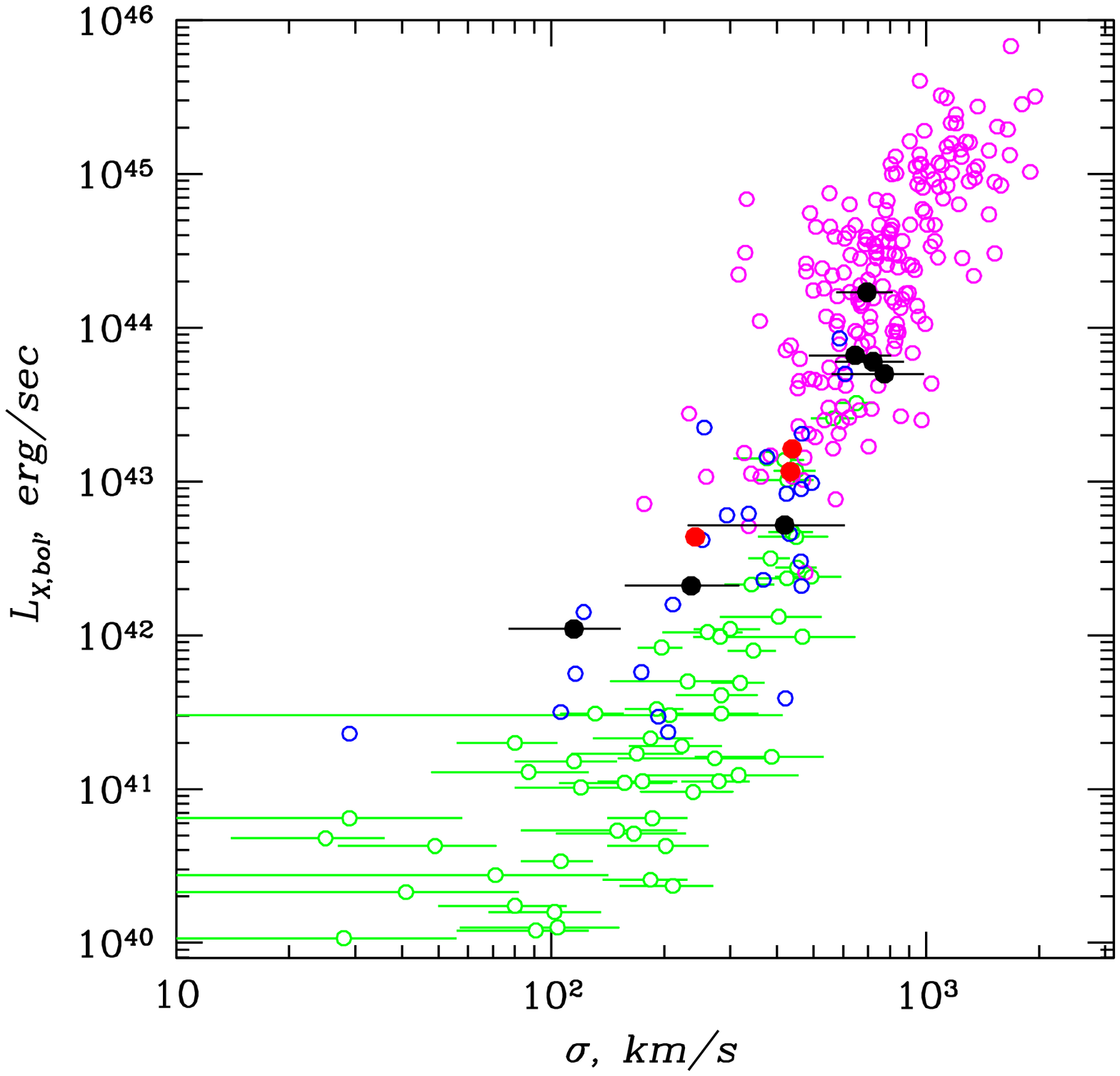}{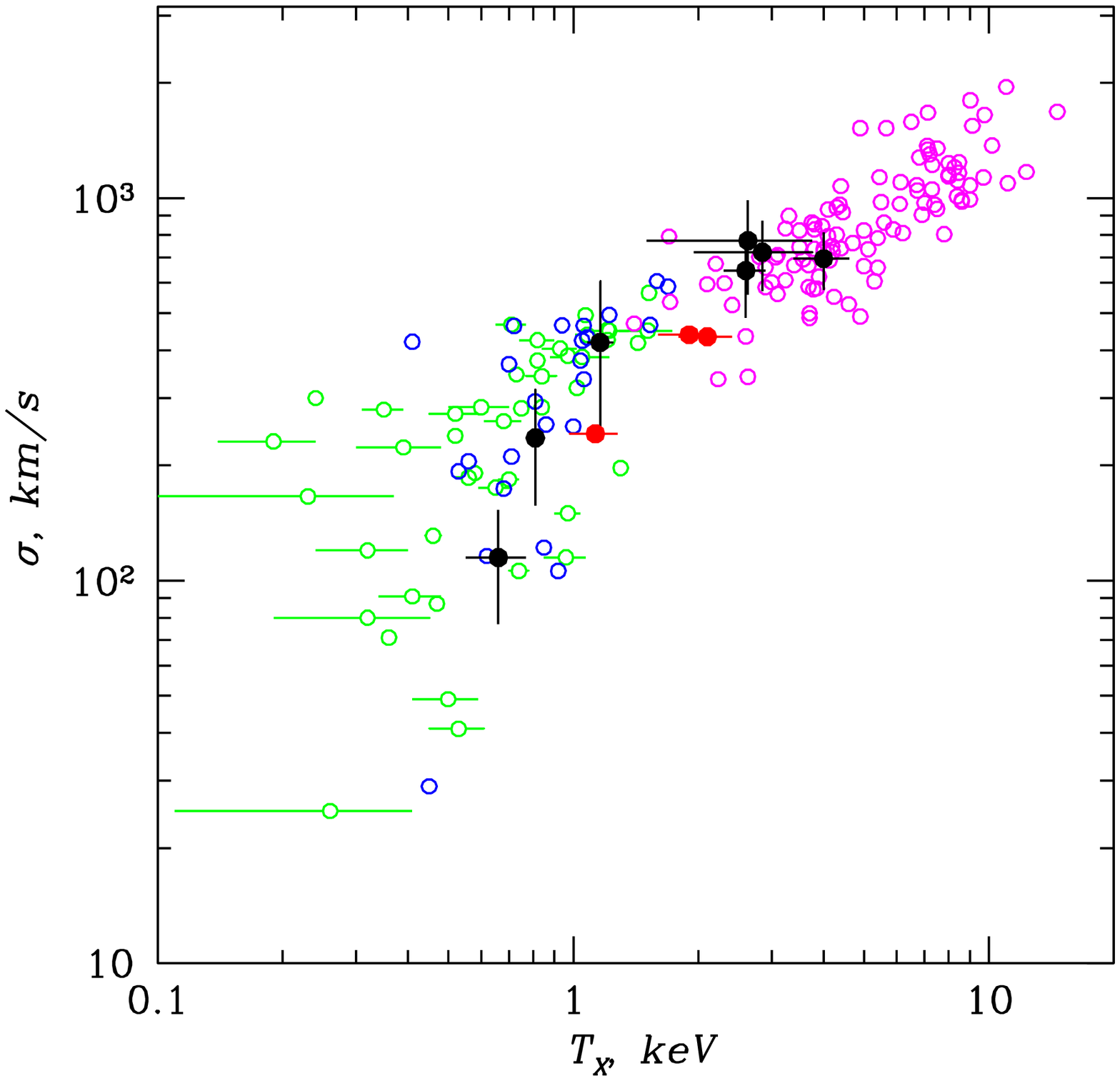}
  \caption{Comparison of the measurements done in this work (red
    points), with the data for fossil groups taken from Khosroshahi et
    al. (2007) (black points). Green points were taken from Osmond \&
    Ponman (2004), blue from Helsdon \& Ponman (2003), and magenta
    from Wu et al. (1999).}
     \label{fig:scale_rel}
\end{figure*}

Hierarchical structure formation theory predicts the formation of
larger galaxies through mergers of smaller galaxies. Studies of the
largest galaxies provide us with a sensitive test for these
predictions. The merger rates depend on masses of merging galaxies and
on the impact parameter at first encounter.  Under the usual
assumption of an isothermal spherical model for each galaxy and a
circular orbit for the smaller halo, the infall time can be estimated
as \cite{2005ApJ...630L.109D}:
\begin{equation}
  t_{inf}=12.4~Gyr
  \left(\frac{r_0}{100~kpc}\right)
  \left(\frac{V_M}{700~km/s}\right)^2\left(\frac{250~km/s}{V_S}\right)^3
\end{equation}
where $r_0$ is an impact parameter, and $V_M$ and $V_S$ are
characteristic circular velocities for the main halo and the
satellite, correspondingly. One conclusion from this formula is that
the mergers are more efficient in groups ($V_M \lesssim$ 700~km/s),
than in rich clusters ($V_M \gtrsim$ 1400 km/s). Another conclusion is
that $L_*$ galaxies with $V_S \sim$ 250~km/s merge faster than dwarf
galaxies. For mergers along filaments with $r_0 \lesssim$ 10~kpc, a
deficit of $L_*$ galaxies is expected within the groups of mass
10$^{13}$~--~10$^{14}M_\odot$ dominated by a single giant elliptical
galaxy. While this model explains the existence and properties of the
observed fossil groups and OLEGs, we would like to point out that
there is no clear physical justification to define a magnitude gap
observed in such systems at some particular level.  In this dynamical
sense, the fossil group criteria discussed in the literature
\cite{2003MNRAS.343..627J} seem to be rather arbitrary. Recent
statistical studies of SDSS galaxy groups \cite{2007arXiv0710.5096Y}
confirm that the $\Delta m_{12}$ magnitude gap distribution is smooth
and exhibits no special features at $\Delta m_{12} \geqslant$ 2.
There is no clear guidance for the choice of other selection criteria,
like the radius of the optical search for galaxies and the acceptable
range of redshifts. In fact, Santos et al. (2007) applied different
cuts to SDSS data and came to lists of 6 to 34 candidates.  Exploring
the parameter space may be more fruitful in this situation than by
setting particular cuts and limits.

The central galaxy of the group rxj1505 is much brighter than the
others in the group, however, the absolute difference in magnitudes
between the first and second brightest galaxy is 1.69, which is not
enough to satisfy the commonly used fossil group
definition~\cite{2003MNRAS.343..627J}.  While this group is not
exceptionally bright or massive, its mass-to-light ratio and other
X-ray scaling relations are similar to fossil groups~\cite
{2007MNRAS.377..595K}.  The magnitude gap for this system is close to
that of the group Cl~1205+44, which at z=0.59 has been identified as
the most distant known fossil group \cite{2005ApJ...624..124U}.

The group rxj0029 appears to be a more unusual object, as in this case
the largest magnitude gap is not between the brightest and the second
brightest galaxy, but between the second brightest and the rest of the
group. While an absence of $L_*$ galaxies and a large magnitude gap
between the two brightest galaxies and the third brightest galaxy
($\Delta m_{13}$=2.47, $\Delta m_{23}$=2.07 --- for galaxies inside
$r_{2500}$) assume an evolved system, the presence of the second
bright galaxy breaks the standard picture.  In the optical the two
brightest galaxies look rather similar, close in their luminosities,
morphologies and colors (see Fig.~\ref{fig:contour_images},\
\ref{fig:cmrs} and Table~\ref{tab:optical_par}). However, the X-ray
emission is clearly centered around the brightest object, with its
peak at the position of the central galaxy
(Fig.~\ref{fig:contour_images}). One possibility is that in spite of
similar redshifts the second brightest galaxy is X-ray faint and
simply projected on the usual fossil group.  However, the weighted
center of the X-ray emission in rxj0029 is shifted from its peak to
the direction of the second brightest galaxy.  This may be an
indication that the two galaxies are interacting.  One may speculate
that this interaction is too recent for the stars of the two galaxies
to merge, but still long enough, so that the gas component is stripped
from the second galaxy and moved toward the center of the group.
Continuing infall of that gas may explain an excess of X-ray emission
toward the second galaxy. Whether this picture is accurate or not, it
demonstrates again that fossil and fossil-like groups may be a more
heterogeneous population than is usually assumed.

The third group, ugc00842, satisfies the standard fossil group
definitions. The magnitude gap between the first and second brightest
galaxy is greater than 2 (see Table~\ref{tab:optical_par},
Fig.~\ref{fig:cmrs}). Moreover, it holds for all member galaxies lying
inside $r_{500}$. The next bright galaxy (01:19:13.46,~$-$01:08:41.2),
for which $m_{12}\approx1$ is 525 kpc away from the central galaxy of
ugc00842. The group is very bright in the X-ray relative to its
optical luminosity, as is typical for fossil groups
(Fig.~\ref{fig:scale_rel}).  Thus, this object can be added to the
list of known fossil groups (see~Mendes de Oliveira et al. 2006).

Observationally, the most notable difference between fossil groups and
the general population of galaxy groups is that fossils are brighter
in the X-ray at the same optical luminosity. Our selection criteria
(described in Section 2.1) provided us with a range of systems
including a fairly average group, an evolved group, and a fossil
group. However, their locations on the optical/X-ray scaling-laws
indicate that all three appear similar to the fossil group sample
studied in Khosroshahi et al.~(2007).

As a consequence, it may be that fossil groups, as classically
defined, do not reside in a preferred location in this parameter
space. Instead, it may be that the currently available group catalogs
fail to accurately trace the full range of parameter space in the
scaling-laws.

Historically, fossil groups are characterized by their unusual X-ray
and optical properties.  While these systems are supposed to be
common, less than 20 of them have been studied so far.  This is
understandable, due to the fact that X-ray observations have so far
been very spotty or shallow, and as a result it has been hard to
collect adequate statistics for these objects.  In contrast, optical
data provided by the SDSS contains hundreds of thousands of galaxy
groups suitable for this type of study. In this work we use optical
selection criteria to identify possible fossil group candidates, which
can then be further studied with dedicated X-ray observations. We
focused on the observation of giant elliptical galaxies, which
dominate a population of dwarf galaxies and are bright in X-rays
because of their significant masses and large amounts of stripped
intergroup gas. We confirmed that in all cases we detect an extended
X-ray emission as expected from groups in the considered mass range.
This allows us to conclude that the algorithm presented in this paper
is quite efficient for the search of groups of galaxies dominated by
giant elliptical galaxy. However, it should be tightened for the
search of fossil groups.

\bigskip

AV thanks Oleg Kotov for useful advice in \emph{XMM-Newton} data
reduction.  We gratefully acknowledge helpful discussions with Alexey
Vikhlinin and Ann Zabludoff. We also acknowledge use of Alexey
Vikhlinin's $zhtools$
software\footnote{http://hea-www.harvard.edu/RD/zhtools/} in our X-ray
data analysis.  We are thankful to Habib Khosroshahi for providing us
with data points for Fig.~\ref{fig:scale_rel}. This work was supported
by the LDRD program at Los Alamos National Laboratory.

\end{document}